\documentclass[pra, aps, twocolumn, showkeys, groupedaddress, showpacs, superscriptaddress]{revtex4-2}
\usepackage{amssymb, amsmath, amsthm, color, graphicx, times, graphicx}
\usepackage[colorlinks={true}]{hyperref}
\hypersetup{colorlinks=true,linkcolor=red,citecolor=blue,urlcolor=blue}
\usepackage{pstricks}
\usepackage{subfigure}
\usepackage{times}
\usepackage{bbold}
\usepackage{color}
\usepackage{comment} 
\usepackage{orcidlink}

\newcommand{\ket}[1]{|{#1}\rangle}

\providecommand{\openone}{\leavevmode\hbox{\small1\kern-4.3pt\normalsize1}}

\theoremstyle{plain}

\theoremstyle{definition}

\begin{document}

\title{Quantumness of hybrid systems under quantum noise}

\author{M. Abdellaoui~\!\!\orcidlink{0009-0002-0801-6042}}\affiliation{LPHE-Modeling and Simulation, Faculty of Sciences, Mohammed V University in Rabat, Rabat, Morocco}
\author{N.-E. Abouelkhir~\!\!\orcidlink{0000-0002-6164-0525}}\affiliation{LPHE-Modeling and Simulation, Faculty of Sciences, Mohammed V University in Rabat, Rabat, Morocco}
\author{A. Slaoui$^*$~\!\!\orcidlink{0000-0002-5284-3240}}\affiliation{LPHE-Modeling and Simulation, Faculty of Sciences, Mohammed V University in Rabat, Rabat, Morocco}\affiliation{Centre of Physics and Mathematics, CPM, Faculty of Sciences, Mohammed V University in Rabat, Rabat, Morocco}
\author{R. Ahl Laamara~\!\!\orcidlink{0000-0001-8410-9983}}\affiliation{LPHE-Modeling and Simulation, Faculty of Sciences, Mohammed V University in Rabat, Rabat, Morocco}\affiliation{Centre of Physics and Mathematics, CPM, Faculty of Sciences, Mohammed V University in Rabat, Rabat, Morocco}\author{S. Haddadi$^*$~\!\!\orcidlink{0000-0002-1596-0763}}\address{School of Particles and Accelerators, Institute for Research in Fundamental Sciences (IPM), P.O. Box 19395-5531, Tehran, Iran\\  $^*${ Corresponding authors:} abdallah.slaoui@um5s.net.ma,  haddadi@ipm.ir}

\begin{abstract}
\textbf{Abstract.} 
We investigate the quantum correlations in an axially symmetric hybrid qubit–qutrit system subjected to different noisy environments. We first introduce a physical model and analyze its Hamiltonian structure, emphasizing the role of hybrid dimensionality and axial symmetry. The effects of decoherence are then examined under two local noise mechanisms, namely dephasing and phase-flip channels, acting on the qubit and qutrit subsystems in both symmetric and asymmetric configurations. Quantum correlations are quantified using negativity to capture entanglement and quantum discord based on linear entropy to characterize more general nonclassical correlations. Our results show that both thermal fluctuations and phase noise lead to a monotonic degradation of quantum correlations, with increasing temperature accelerating coherence loss and inducing entanglement sudden death at finite temperatures. While negativity vanishes abruptly under sufficiently strong noise, quantum discord persists beyond the entanglement threshold, revealing residual quantum correlations in mixed states. We further demonstrate that asymmetric noise configurations significantly enhance the robustness of both entanglement and discord by partially shielding coherence in the less affected subsystem. A comparative analysis reveals that phase-flip noise is more destructive than pure dephasing, leading to faster suppression of quantum correlations. 
\end{abstract}



\maketitle
\section{Introduction}
\vspace{-0.2cm}
Quantum information theory has emerged as one of the most profound developments in modern physics, bridging the gap between information science and quantum mechanics. The foundations of this field trace back to the pioneering work of Einstein, Podolsky, and Rosen, who introduced the concept of quantum entanglement as a manifestation of the nonlocal nature of quantum mechanics~\cite{EPR1935}. Schrödinger further emphasized the central role of entanglement as the characteristic trait of quantum mechanics~\cite{Schrodinger1935}. Over the following decades, entanglement has become recognized as a fundamental resource for numerous quantum information protocols~\cite{Brodutch2011}, including quantum teleportation, superdense coding, and quantum key distribution~\cite{Bennett1993, Ekert1991}. 

However, it was later realized that entanglement does not encompass all forms of quantum correlations, as certain separable states can still exhibit nonclassical features~\cite{Adesso2010}. To move beyond this framework and address the limitation of entanglement as an exclusive resource, Ollivier and Zurek~\cite{Ollivier2001} introduced the concept of quantum discord (QD) as a more general measure capable of quantifying all nonclassical correlations, even in separable states~\cite{Henderson2001}. This measure provides a more complete description of quantum resources and has gained considerable attention for its robustness against decoherence and its operational relevance in quantum computation models where entanglement can vanish~\cite{Datta2008}. Nevertheless, since the analytical computation of QD based on von Neumann entropy is often intractable~\cite{Luo2008, Ali2010, Girolami2011}, alternative approaches have been developed, with the use of linear entropy proving particularly fruitful as it enables closed analytical expressions for QD~\cite{Fanchini2010}, thereby simplifying calculations in higher-dimensional systems such as qubit-qudit models~\cite{Yurischev2025} and facilitating a deeper analysis of the dynamics of quantum correlations under realistic conditions~\cite{Vinjanampathy2012}.

{More broadly, quantum correlations are regarded as the essential resource responsible for the advantages of quantum information processing over classical approaches, providing enhanced security in quantum communication and superior capabilities in quantum computation, as widely recognized in modern quantum information research \cite{Xie2022, Lu2025}. In practical quantum communication protocols, these advantages rely explicitly on the generation and preservation of nonclassical correlations through two-photon interference, Bell-state measurements, or multipartite entangled states, whose successful operation critically depends on phase stability and coherence. Nevertheless, these nonclassical correlations remain highly sensitive to environmental influences. Under realistic conditions, unavoidable interactions with the environment generate quantum noise, such as channel fluctuations, phase drift, and decoherence, which is a dominant mechanism for degrading and possibly erasing quantum correlations. Indeed, maintaining interference visibility and phase coherence constitutes one of the main experimental challenges in realistic quantum communication networks. Investigating the influence of various noise mechanisms on these correlations is therefore crucial for the reliable realization of quantum information protocols.}

The quantum correlations in mixed qubit-qudit ($2 \otimes d$) axially symmetric systems constitute a central area of study in quantum information and quantum metrology due to their distinct mathematical properties and physical relevance~\cite{Yurischev2025}. This particular structure is exploited in numerous quantum applications~\cite{Yurischev2025Sci}. Moreover, several physical systems including nanomagnets, cold atoms in optical traps, spin ensembles, superconducting qubits, or certain nuclear spin systems, naturally exhibit these axial symmetry features, reinforcing the interest in their study~\cite{HagiwaraandKatsumata1999, ABDELLAOUI2024, abouelkhir2025, CencarikovaandNavrátil2020, abouelkhir2025(2)}. 

This work focuses on a hybrid system composed of a spin-$1/2$ (qubit) coupled to a spin-$1$ (qutrit), governed by a general Hamiltonian that preserves axially symmetric and incorporates ten distinct physical parameters describing a wide range of interactions, including external magnetic fields, Heisenberg exchange constants, single- and two-ion anisotropies, the Dzyaloshinsky–Moriya (DM) interaction, and higher-order symmetric and asymmetric spin-coupling terms~\cite{Yurischev2025Sci}. This general model encompasses several physically important cases, including the Zeeman effect, the mixed-spin Heisenberg model relevant for hyperfine interactions, spin-orbit coupling contributions, spin-squeezing, and the antisymmetric DM exchange, which is crucial for chiral magnetic structures and spintronic applications. Finally, by considering the thermal evolution and different decoherence channels (dephasing and phase-flip), we use the negativity to characterize quantum entanglement and linear-entropy-based QD to quantify total quantum correlations, analyzing the dynamics and robustness of these correlations under realistic environmental interactions, thus providing a comprehensive understanding of the interplay between decoherence, temperature, and the preservation of quantum resources in hybrid spin systems.


\section{Axially symmetric states} \label{Sec 2}

Let us consider a hybrid system composed of a qubit coupled to a qutrit, characterized by axially symmetric. The general form of Hamiltonian is~\cite{Yurischev2025Sci}
\begin{equation}
\begin{aligned}\label{eq:19}
\mathcal{H} &= B_1 s_z + B_2 S_z + J(s_x S_x + s_y S_y) + J_z s_z S_z + K S_z^2 \\
& \hspace{0.5cm} +K_1 (S_x^2 + S_y^2) + K_2 s_z S_z^2 + \Gamma [ s_y (S_y S_z + S_z S_y)] \\
&\hspace{0.2cm}  \quad + D_z (s_x S_y - s_y S_x) + \Gamma [s_x (S_x S_z + S_z S_x) ] \\
 &\hspace{0.2cm}\quad + \Lambda [s_x (S_y S_z + S_z S_y) - s_y (S_x S_z + S_z S_x)],
\end{aligned}
\end{equation}
In this expression, $s_i = \sigma_i / 2$, where $\sigma_i$ are the Pauli matrices that represent the spin-$1/2$ operators associated with the qubit, while $S_i$ denotes the spin-1 matrices corresponding to the qutrit. Moreover, $ B_1 $ and $ B_2 $ represent the $ z$-components of external magnetic fields acting on the qubit and qutrit, respectively; $ J $ and $ J_z $ are the Heisenberg exchange constants; $ K $ and $ K_1 $ denote the uniaxial and planar single-ion anisotropies; $ K_2 $ stands for the two-ion uniaxial anisotropy; $ D_z $ is the $ z$-component of the DM vector; and $ \Gamma $ and $ \Lambda $ represent symmetric and asymmetric higher-order spin-coupling terms newly introduced by Yurischev et al.~\cite{Yurischev2025Sci}.

\subsection{System at thermal equilibrium}
At thermal equilibrium, the state of a general system is described by the Gibbs density operator, defined as
\begin{equation}
\rho = \frac{1}{Z} \exp(-\mathcal{H}/T).
\label{eq:27}
\end{equation}
For our considered Hamiltonian~\eqref{eq:19}, the Gibbs density matrix is given by
\begin{equation}
    \rho_T = \begin{pmatrix}
\rho_{11} & 0 & 0 & 0 & 0 & 0 \\
0 & \rho_{22} & 0 & \rho_{24} & 0 & 0 \\
0 & 0 & \rho_{33} & 0 & \rho_{35} & 0 \\
0 & \rho_{24}^{*} & 0 & \rho_{44} & 0 & 0 \\
0 & 0 & \rho_{35}^{*} & 0 & \rho_{55} & 0 \\
0 & 0 & 0 & 0 & 0 & \rho_{66}
\end{pmatrix},
\label{H zero}
\end{equation}
where 
\begin{equation}
\begin{aligned}
\rho_{11} &= \frac{1}{Z} e^{-E_1/T},\quad \rho_{66} = \frac{1}{Z} e^{-E_6/T}, \\
\rho_{22} &= \frac{1}{Z} \left( \cosh \frac{r_1}{2T} + \frac{h_3 - h_1}{r_1} \sinh \frac{r_1}{2T} \right) e^{-(h_1 + h_3)/2T}, \\
\rho_{33} &= \frac{1}{Z} \left( \cosh \frac{r_2}{2T} + \frac{h_4 - h_2}{r_2} \sinh \frac{r_2}{2T} \right) e^{-(h_2 + h_4)/2T}, \\
\rho_{44} &= \frac{1}{Z} \left( \cosh \frac{r_1}{2T} + \frac{h_1 - h_3}{r_1} \sinh \frac{r_1}{2T} \right) e^{-(h_1 + h_3)/2T}, \\
\rho_{55} &= \frac{1}{Z} \left( \cosh \frac{r_2}{2T} + \frac{h_2 - h_4}{r_2} \sinh \frac{r_2}{2T} \right) e^{-(h_2 + h_4)/2T}, \\
\rho_{24} &= -\frac{2g_1}{Zr_1} \sinh \frac{r_1}{2T} e^{-(h_1 + h_3)/2T}, \\
\rho_{35} &= -\frac{2g_2}{Zr_2} \sinh \frac{r_2}{2T} e^{-(h_2 + h_4)/2T},\nonumber
\end{aligned}
\label{eq:28}
\end{equation}
with
\begin{equation}
\begin{aligned}
&E_{1,6} = J_z/2 + K + K_1 \pm (B_1/2 + B_2 + K_2/2),\\
&h_1 = B_1/2 + 2K_1,\quad h_4 = -B_1/2 + 2K_1,\\
&h_{2,3} = \pm B_1/2 \mp B_2 - J_z/2 + K + K_1 \pm K_2/2, \\
&g_{1,2} = [J \pm \Gamma + i(D_z \pm \Lambda)]/\sqrt{2},\nonumber
\end{aligned}
\label{eq:22}
\end{equation}
\begin{align}
r_1 = \sqrt{(h_1 - h_3)^2 + 4|g_1|^2}, \nonumber\\ r_2 = \sqrt{(h_2 - h_4)^2 + 4|g_2|^2},\nonumber
\label{eq:24}
\end{align}
and the partition function is
\begin{equation}
\begin{aligned}
Z &= 2 \left[ \cosh \frac{B_1 + 2B_2 + K_2}{2T} e^{-(J_z + 2K+2K_1)/2T} \right. \\
&\quad \left. + \cosh \frac{r_1}{2T} e^{-(h_1 + h_3)/2T} + \cosh \frac{r_2}{2T} e^{-(h_2 + h_4)/2T} \right].\nonumber
\end{aligned}
\label{eq:29}
\end{equation}

\subsection{Thermal state under decoherence}
\label{Sec 3}
In realistic quantum systems, it is essential to consider the possible degradation of any initially prepared entanglement due to the inevitable interaction with the surrounding environment, commonly referred to as decoherence \cite{Gaidi2026}. In this subsection, we analyze the dynamics of a hybrid qubit-qutrit system subjected to typical noise channels, namely the dephasing and phase-flip channels. To explore the influence of environmental interactions, we examine two distinct decoherence scenarios. { In the first case, we consider symmetric decoherence, where both subsystems (qubit and qutrit) are equally affected by the environment; in the second, only the qubit is exposed to decoherence. In the latter case, we focus on the physically relevant scenario where the decoherence acts independently on the qubit subsystem, while the qutrit remains unaffected. This corresponds to a situation in which the environment interacts asymmetrically with the hybrid system.}

To model the influence of these processes, we employ the Kraus operator formalism, which effectively describes the non-unitary evolution of open quantum systems.
{In general, the time evolution of the density matrix of a bipartite system under local quantum noise can be written in the Kraus representation as
\begin{equation}
\rho^{AB} (t) = \sum_{i,j} K_{ij} \, \rho^{AB}(0) \, K_{ij}^\dagger,
\end{equation}
where the Kraus operators $K_{ij}$ satisfy the completeness relation $\sum_{i,j} K_{ij}^\dagger K_{ij} = I$. The explicit form of these operators depends on the physical nature of the noise channel acting on each subsystem.}
Specifically, the local dephasing channel responsible for the gradual decay of quantum coherence without energy exchange can be represented by a set of Kraus operators acting on each subsystem. Accordingly, the time-dependent density operator of the qubit-qutrit system, initially prepared in the thermal state $\rho^{AB}(0)=\rho_T$, given in Eq.~\eqref{H zero}, evolves as 
\begin{equation}
\rho^{AB} (t) = \sum_{i=1}^{2}\sum_{j=1}^{3} F_{j}^{B} E_{i}^{A} \rho^{AB}(0) E_{i}^{A\dagger} F_{j}^{B\dagger}.
\label{eq14}
\end{equation}
{ where $E_{i}^{A}$ and $F_{j}^{B}$ are the Kraus operators related to the qubit and qutrit subsystems, respectively.
}

{
In this work, we model the asymmetric environmental influence as acting solely on the qubit subsystem. This choice is motivated by both experimental relevance and theoretical clarity: such a situation naturally occurs in hybrid platforms where a two-level system is more directly exposed to environmental fluctuations than a higher-dimensional one, particularly in electron–nuclear spin composites and circuit-QED architectures~\cite{Jelezko2004, Maurer2012,Haddadiscirep2026}. This modeling strategy has been adopted in previous investigations of qubit–qutrit systems under decoherence, where asymmetric couplings are used to highlight dimension-dependent effects on quantum correlations~\cite{Karpat2011, Cardenas2017,Manan2025}. The formalism we present can also accommodate noise acting on the qutrit or both subsystems, but this would not qualitatively change the main insights regarding the role of subsystem dimensionality on correlation robustness.}

The set of Kraus operators for a single qubit $A$ and a single qutrit $B$ that reproduce the effect of a dephasing channel are given by
\begin{align}
E_{1}^{A} &= \mathrm{diag}\!\left(1,\sqrt{1-\gamma_{A}(t)}\right) \otimes I_{3},\nonumber\\
E_{2}^{A} &= \mathrm{diag}\!\left(0,\sqrt{\gamma_{A}(t)}\right) \otimes I_{3},\nonumber\\
F_{1}^{B} &= I_{2} \otimes \mathrm{diag}\!\left(1,\sqrt{1-\gamma_{B}(t)},\sqrt{1-\gamma_{B}(t)}\right), \\[4pt]
F_{2}^{B} &= I_{2} \otimes \mathrm{diag}\!\left(0,\sqrt{\gamma_{B}(t)},0\right), \nonumber\\[4pt]
F_{3}^{B} &= I_{2} \otimes \mathrm{diag}\!\left(0,0,\sqrt{\gamma_{B}(t)}\right).\nonumber
\end{align}
The time-dependent parameters $\gamma_A(t)$ and $\gamma_B(t)$ characterize the decoherence dynamics of the system and are defined by the exponential decay functions
\begin{equation}
\gamma_{A}(t) = 1 - e^{-t\Gamma_{A}} \quad \text{and} \quad \gamma_{B}(t) = 1 - e^{-t\Gamma_{B}},
\end{equation}
where $t \geq 0$ represents the time variable, and $\Gamma_{A} > 0$ and $\Gamma_{B} > 0$ denote the decay rates associated with subsystems $A$ and $B$, respectively. These parameters satisfy the following constraints
\begin{equation}
\gamma_A(t), \gamma_B(t) \in \mathbb{R}, \,\, \text{with} \,\,\, 0 \leq \gamma_A(t) \leq 1 \,\,\, \text{and} \,\,\, 0 \leq \gamma_B(t) \leq 1.
\end{equation}
At $t = 0$, both parameters vanish ($\gamma_A(0) = \gamma_B(0) = 0$), indicating no decoherence, while as $t \to \infty$, they asymptotically approach unity ($\gamma_A(\infty) = \gamma_B(\infty) = 1$), corresponding to complete decoherence. The quantities $\Gamma_A$ and $\Gamma_B$ determine the characteristic time scales of the decoherence processes, with larger values leading to faster convergence to the steady state.\par
After some algebraic calculations, the matrix form of time-evolved state~\eqref{eq14} for a dephasing channel is obtained as
\begin{equation} 
\rho_{\text{Dc}}^{AB} (t)= \begin{pmatrix}
\rho_{11} & 0 & 0 & 0 & 0 & 0 \\
0 & \rho_{22} & 0 & k \rho_{24} & 0 & 0 \\
0 & 0 & \rho_{33} & 0 & l \rho_{35} & 0 \\
0 & k\rho_{24}^{*} & 0 & \rho_{44} & 0 & 0 \\
0 & 0 & l\rho_{35}^{*} & 0 & \rho_{55} & 0 \\
0 & 0 & 0 & 0 & 0 & \rho_{66}
\end{pmatrix},
\label{eq21}
\end{equation}
with $k = \sqrt{(\gamma_A(t) - 1)(\gamma_B(t) - 1)}$ and $l = (1-\gamma_B(t) ) \sqrt{\gamma_A(t)-1}$.


The Kraus operators describing the phase-flip channel for a single qubit $A$ are given by
\begin{align}
E_{1}^{A} &= \sqrt{1-\frac{\gamma_{A}(t)}{2}}
\mathrm{diag}\!\left(1,1\right)
\otimes I_{3}, \\
E_{2}^{A} &= \sqrt{\frac{\gamma_{A}(t)}{2}}
\mathrm{diag}\!\left(1,-1\right)
\otimes I_{3},
\end{align}
and those for a single qutrit $B$ can be written as
\begin{align}
F_{1}^{B} &= I_{2} \otimes \sqrt{1-\frac{2\gamma_{B}(t)}{3}}
\mathrm{diag}\!\left(1,1,1\right), \\
F_{2}^{B} &= I_{2} \otimes \sqrt{\frac{\gamma_{B}(t)}{3}}
\mathrm{diag}\!\left(1,e^{-i2\pi/3},e^{i2\pi/3}\right), \\
F_{3}^{B} &= I_{2} \otimes \sqrt{\frac{\gamma_{B}(t)}{3}}
\mathrm{diag}\!\left(1,e^{i2\pi/3},e^{-i2\pi/3}\right),
\end{align}
Similar to the dephasing channel, 
one can obtain the time-evolved density-matrix 
under a phase-flip channel as follows
\begin{equation}
\rho_{\text{Pc}}^{AB} (t) = 
\begin{pmatrix}
\rho_{11} & 0 & 0 & 0 & 0 & 0 \\
0 & \rho_{22} & 0 & k^2 \rho_{24} & 0 & 0 \\
0 & 0 & \rho_{33} & 0 & k^2 \rho_{35} & 0 \\
0 & k^2\rho_{24}^{*} & 0 & \rho_{44} & 0 & 0 \\
0 & 0 & k^2\rho_{35}^{*} & 0 & \rho_{55} & 0 \\
0 & 0 & 0 & 0 & 0 & \rho_{66}
\end{pmatrix}.
\label{eq28}
\end{equation}

\section {Dynamics of Entanglement in the Qubit–Qutrit System} \label{Sec 4}
The emergence of quantum information as a discipline has placed entanglement at the heart of resources exploitable for computation and communication. Since 1990, research has intensified to establish operational criteria for distinguishing separable states from entangled states, particularly in the case of mixed states. It was in this context that the positive partial transpose (PPT) criterion was proposed by Peres \cite{Peres1996} and then rigorously established by  Horodecki family \cite{Horodecki1996}. This criterion, although necessary and sufficient for separability in low-dimensional systems, did not in itself provide a quantification of entanglement. To fill this gap, Vidal and Werner \cite{Vidal2002} introduced the measure now known as negativity. It offers a direct quantitative translation of the violation of the PPT criterion, defined from the sum of the negative eigenvalues of the partially transposed density matrix~\cite{Mansour2021}.

Negativity thus presents itself as both a natural and computable measure of entanglement. Formally, for a bipartite state $\rho^{AB}$ in a system of dimension $d \otimes d'$, it is expressed by 
\begin{equation}
N(\rho^{AB}) = \frac{\|\rho^{T_s}\|_1 - 1}{d - 1},
\end{equation}
where $\| \cdot \|_1$ denotes the trace norm. This expression concisely captures the degree of non-positivity of the partial transpose, providing an operational estimate of the entanglement contained in the state. Although it has the drawback of vanishing for PPT entangled states, so-called bound entangled states, its broad applicability and ease of computation make it a privileged tool for the analysis of bipartite systems, whether dealing with qubit-qudit pairs or more general devices encountered in quantum metrology and quantum information processing protocols. Negativity remains, in this regard, a cornerstone of entanglement metrics in the theoretical and applied landscape of quantum information.\par
For an arbitrary mixed state $\rho^{AB}$ in a $2 \otimes 2$ or a $2 \otimes 3$ system, the negativity can well be characterized and quantified by  \cite{Vidal2002, Lee2003, Abdellaoui2026}
\begin{equation}
N(\rho^{AB}) = \|\rho_{AB}^{T_B}\|_1 - 1,
\end{equation}
which represents the absolute value of the sum of the negative eigenvalues of the partially transposed density matrix $\rho_{AB}^{T_B}$. 
The partial transpose $\rho_{AB}^{T_B}$, taken with respect to subsystem $B$ in an arbitrary product orthonormal basis 
$\{ f_i \otimes f_j \}$, is defined through its matrix elements as
\begin{equation}
(\rho_{AB}^{T_B})_{m\mu,n\nu} \equiv 
\langle f_m \otimes f_\mu | \rho_{AB}^{T_B} | f_n \otimes f_\nu \rangle 
= \rho_{m\nu,n\mu}.
\end{equation}
Hence, one can write
\begin{equation}
N(\rho^{AB}) = 2 \max\{0, -\lambda_S\},
\end{equation}
where $\lambda_S$ represents the sum of all negative eigenvalues of $\rho_{AB}^{T_B}$. 

After recalling these general aspects of negativity, we now investigate its behavior under the action of the two decoherence channels introduced in the previous section, namely the dephasing and the phase-flip channels.

In the first case, each subsystem is subjected to a dephasing channel. This type of noise acts only on the off-diagonal elements of the density matrix, leading to a gradual loss of coherence without any exchange of energy with the environment. The analysis of this channel allows us to highlight the role of pure dephasing in the gradual degradation of entanglement and the possible emergence of the sudden death phenomenon.
\\
To obtain an analytical expression for negativity, we first form the partial transpose matrix $\rho_{\text{Dc}}^{T_B}$ of our state \eqref{eq21} as
\begin{equation}
    \rho_{\text{Dc}}^{T_B} = \begin{pmatrix}
\rho_{11} & 0 & 0 & 0 & k  \rho_{24} & 0 \\
0 & \rho_{22} & 0 & 0 & 0 & l  \rho_{35} \\
0 & 0 & \rho_{33} & 0 & 0 & 0 \\
0 & 0 & 0 & \rho_{44} & 0 & 0 \\
k  \rho_{24}^{*} & 0 & 0 & 0 & \rho_{55}& 0 \\
0 & l \rho_{35}^{*} & 0 & 0 & 0 & \rho_{66}
\end{pmatrix}.
\label{eq22}
\end{equation}
Next, the possible negative eigenvalues of the above matrix \eqref{eq22} can be expressed in general as
\begin{align}
\lambda_{2}^{\text{Dc}} &= \frac{\rho_{11} + \rho_{55} 
- \sqrt{(\rho_{11} - \rho_{55})^{2} + 4\,|\rho_{24}|^{2}\,k^{2}}}{2}, \nonumber\\
\lambda_{4}^{\text{Dc}} &= \frac{\rho_{22} + \rho_{66} 
- \sqrt{(\rho_{22} - \rho_{66} )^{2} + 4\,|\rho_{35}|^{2}\,l^{2}}}{2}.
\end{align}
Finally, the corresponding negativity is 
\begin{equation}
N(\rho_{\text{Dc}}) = 2 \, \max \big(0,\; -\lambda_{2}^{\text{Dc}} - \lambda_{4}^{\text{Dc}} \big).
\end{equation}


In the second case, we consider a phase-flip channel, where the phase of the basis states is inverted with a certain probability. Unlike the dephasing channel, this type of noise introduces sign inversions in the phase elements, altering the correlation structure between the subsystems. 

The comparative study between these two channels reveals distinct decoherence dynamics and provides deeper insight into the robustness of quantum correlations against different types of dissipative processes.


{
Starting from the evolved density matrix given in Eq.~\eqref{eq28}, and following the same procedure as for the dephasing case, the potentially negative eigenvalues are obtained as follows:
\begin{equation}
\lambda_2^{\text{Pc}} =
\frac{\rho_{11} + \rho_{55} - \sqrt{(\rho_{11} - \rho_{55})^2 + 4 \left| \rho_{24} \right|^2 k^4}}{2},
\end{equation}

\begin{equation}
\lambda_4^{\text{Pc}} =
\frac{\rho_{22} + \rho_{66} - \sqrt{(\rho_{22} - \rho_{66})^2 + 4 \left| \rho_{35} \right|^2 k^4}}{2},
\end{equation}
Thus, the negativity under the phase-flip channel is given by
\begin{equation}
N(\rho_{\text{Pc}}) = 2 \max \left( 0, -\lambda_2^{\text{Pc}} - \lambda_4^{\text{Pc}} \right).
\end{equation}

}

\section {Quantum Discord based on Linear Entropy in Hybrid Systems } \label{Sec 5}
Quantum discord based on linear entropy is a measure of quantum correlations beyond entanglement~\cite{Ma2015, Ali2020, Abouelkhir2026}, highlighting the nonclassical aspects of quantum systems. Unlike von Neumann entropy, linear entropy provides a simpler and more efficient approach to quantifying the loss of information when a subsystem is measured, making it particularly suitable for mixed states where entanglement alone fails to describe all quantum correlations~\cite{Slaoui2019}. More generally, QD arises from the difference between two classically equivalent definitions of correlations, which no longer coincide in quantum systems due to the non-commutative nature of quantum measurements~\cite{Luo2008}. This concept has been extensively studied over the past twenty years in various contexts~\cite{Ali2020}, particularly for characterizing correlations in Hilbert spaces of different dimensions~\cite{Osborne2006}.
The analytical expression of quantum discord based on linear entropy~\cite{Ali2020, Abouelkhir2024} is obtained for any qubit-qudit system as

\begin{equation} \label{Q General}
Q(\rho^{AB}) = I(\rho^{AB}) - J_2(\rho^{AB}),
\end{equation}
where $I(\rho^{AB})$ is the quantum mutual information defined as
\begin{equation} \label{QMI}
I(\rho^{AB}) = S(\rho^A) + S(\rho^B) - S(\rho^{AB}),
\end{equation}
If the density matrix $\rho$ has eigenvalues $\lambda_i$ such that $\lambda_i \ge 0$ and $\sum_i \lambda_i = 1$, then the von Neumann entropy can be written as  
$S(\rho) = - \sum_i \lambda_i \log \lambda_i$.  
For a pure state, only one eigenvalue equals $1$, and consequently $S(\rho) = 0$. In contrast, for a maximally mixed state in a $d$-dimensional Hilbert space, where $\rho = I/d$, the entropy reaches its maximum value $S(\rho) = \log d$.  Also, for a diagonal qubit state of the form $\mathrm{diag}(p,~1-p)$, the entropy is given by $S(\rho) = -p \log p - (1-p)\log(1-p)$.\par
To calculate the quantum discord $Q(\rho^{AB})$, we proceed by determining the classical correlations of the bipartite quantum system, denoted by $J_2(\rho_{AB})$, which is then combined with the previously obtained mutual information $I(\rho^{AB})$. However, for a qubit-qutrit system, the decomposition of a mixed state $\rho^A \otimes \rho^B$ is given by the Fano-Bloch representation as
\begin{align}
\rho^{AB} =& \frac{1}{6} \bigg(R_{00}~I_2 \otimes I_3 + \sum_{\alpha=1}^{3} R_{\alpha 0} \, \sigma^{\alpha} \otimes I_3  \nonumber\\
&+ \sum_{\beta=1}^{8} R_{0\beta }\, I_2 \otimes \gamma^{\beta} + \sum_{\alpha=1}^{3} \sum_{\beta=1}^{8} R_{\alpha\beta} \, \sigma^{\alpha} \otimes \gamma^{\beta} \bigg),
\end{align}
where $\sigma^{\alpha}$ ($\alpha=1,2,3$) are the Pauli operators acting on the qubit subspace, and $\gamma^{\beta}$ ($\beta=1,\dots,8$) are the Gell-Mann matrices acting on the qutrit subspace. Here, by tracing out the qutrit subsystem $B$, one can obtain the reduced density matrix that describes the local states of the qubit subsystem $A$, given by 
\begin{equation}
   \rho^A = \text{Tr}_B(\rho^{AB}) =\frac{1}{2} \left(  I_2 + \sum_{\alpha=1}^{3} R_{\alpha 0} \sigma^{\alpha} \right). 
\end{equation}
Similarly, tracing out the qubit subsystem $A$ yields the reduced density matrix of the qutrit subsystem $B$, expressed as
\begin{equation}
\rho^B = \text{Tr}_A(\rho^{AB}) =\frac{1}{3} \left(  I_3 + \sum_{\beta=1}^{8} R_{0\beta} \gamma^{\beta} \right),
\end{equation}
where the coefficients $R_{\alpha\beta}$ represent the components of the total correlation tensor associated with $\rho^{AB}$, defined as
\begin{equation} \label{R}
    R_{\alpha\beta} = \operatorname{Tr}[\rho^{AB} \, \sigma^{\alpha} \otimes \gamma^{\beta}].
\end{equation}
In this work, we adopt an alternative method to determine these elements in an arbitrary $2 \otimes d$ system, where we set $d=3$ to describe a qubit–qutrit system. To achieve this, we begin by expressing the qubit-qudit density matrix $\rho^{AB}$ as
\begin{equation} 
\rho^{AB} = \frac{1}{2d} \sum_{\alpha=0}^{3} \sum_{\beta=0}^{d^2-1} R_{\alpha\beta} \sigma^{\alpha} \otimes \gamma^{\beta}.
\end{equation}
A qudit state can be represented in the Bloch form as \cite{Ma2015}
\begin{equation} 
\rho = \frac{I_{d} + {r} \cdot {y}}{d},
\end{equation}
where \(I_{d}\) is the \(d \times d\) identity matrix, \({r}\) is a real vector of dimension \((d^{2} - 1)\), and \(y = (\gamma_{1}, \gamma_{2}, \ldots, \gamma_{d^{2}-1})^{T}\) is the vector whose components correspond to the generators of SU\((d)\). 
The bipartite quantum state \(\rho^{AB}\) can be expressed in the form
\begin{equation} \label{vecr AA}
\rho^{AB} = I_{A} \left( |v^{A^{\prime}A}\rangle \langle v^{A^{\prime}A}| \right)\otimes \Lambda,
\end{equation}
where \(|v^{A^{\prime}A}\rangle\) represents the symmetric purification of the reduced density operator \(\rho^{A}\) with an auxiliary system \(A^{\prime}\). 
Here, \(\Lambda\) is a qudit channel (a completely positive trace-preserving map) that maps a single-qubit state from system \(A^{\prime}\) to a qudit state in system \(B\). The state \(\rho^{A^{\prime}A}\) can be written as
\begin{equation} \label{stat AA}
\rho^{A^{\prime}A} = \frac{1}{4} \sum_{\alpha,\beta=0}^{3} {r}_{\alpha\beta} \sigma^{\alpha} \otimes \sigma^{\beta},
\end{equation}
where \({r}_{\alpha\beta}\) stands for the components of the total correlation tensor associated with \(\rho^{A^{\prime}A}\), given by
\begin{equation} \label{R{'}}
{r}_{\alpha\beta} = \text{Tr}\left(\rho^{A^{\prime}A} \sigma^{\alpha} \otimes \sigma^{\beta}\right).
\end{equation}
The qudit channel $\Lambda$ introduced in Eq.~\eqref{vecr AA} can equivalently be expressed as
\begin{equation} \label{Qudit channel A}
\Lambda(\sigma^{\alpha}) = \sum_{\alpha=0}^{3} \sum_{k=0}^{d^2-1} \mathcal{M}_{\alpha k} \gamma^{k},
\end{equation}
where
\begin{equation} \label{L}
\mathcal{M}_{\alpha \delta} = \text{Tr}\left(\Lambda(\sigma^{\alpha}) \cdot \gamma^{\delta}\right).
\end{equation}
Therefore, by employing Eqs.~\eqref{stat AA} and \eqref{Qudit channel A}, one gets
\begin{equation}
(I \otimes \Lambda) \rho^{A^{\prime}A} = \frac{1}{4} \sum_{\alpha=0}^{3} \sum_{k=0}^{d^2-1} \left({r}\mathcal{M}\right)_{\alpha k} \sigma^{\alpha} \otimes \gamma^{k}.
\end{equation}
Using now Eq.~\eqref{vecr AA}, the matrix \(\mathcal{M}\) is written as
\begin{equation}
\mathcal{M} = \frac{2}{d} ({r}^{-1}) R.
\end{equation}
Let $R, \; r, \; \text{and} \; \mathcal{M}$ are three matrices whose elements are defined by \( R_{\alpha\beta} \) \eqref{R}, \( r_{\alpha\beta} \) \eqref{R{'}} and \( \mathcal{M}_{\alpha\beta} \) \eqref{L}, respectively. Based on these definitions, the real matrix introduced in Eq.~\eqref{Jl} is $M \in \mathbb{R}^{(d^{2}-1) \times 3},$ where each row corresponds to one of the generators of the SU($d$) algebra, and each column represents the three spatial components of the qubit subsystem. The analytical expressions of the elements of $M$ are derived according to the procedure established in Ref.~\cite{Slaoui2020}, namely
\begin{equation}
M_{ij} = \frac{1}{d} \sum_{i=1}^{d^{2}-1} \sum_{j=1}^{3} \mathrm{Tr} \left( \Lambda(\sigma_{i}) \gamma_{j} \right).
\end{equation}
The matrix \( M \) defines the transformation of the Bloch vector \( {r_A} \) from the single-qubit space to the qudit space under the influence of the channel \( \Lambda \), which is written as
\begin{equation} \label{LLL}
M = \sum_{i=1}^{d^{2}-1} \sum_{j=1}^{3} \frac{2}{d} ({r}^{-1} R)_{ij}.
\end{equation}
After determining the eigenvalues of the matrix $(M^{T}M)$, one can identify the maximum eigenvalue $\lambda_{\max}(M^{T}M)$, which plays a crucial role in quantifying the classical correlations of the bipartite quantum system. Hence, we obtain
\begin{equation} \label{Jl}
J_{2}(\rho^{AB}) = \frac{d^{2}}{4} \lambda_{\max}(M^{T}M) \mathcal{S}_{2}(\rho^{A}),
\end{equation}
where the linear entropy $ \mathcal{S}_{2}(\rho^{A}) $ is defined as
\begin{equation}
     \mathcal{S}_{2}(\rho^{A}) = 2(1 - \text{Tr}(\rho_{AB}^2)).
\end{equation}
Then, the analytical form of the classical correlations for any arbitrary $2 \otimes d$ quantum state depends explicitly on this maximal eigenvalue and on the linear entropy of the reduced density matrix $\rho^{A}$.

For the channel corresponding to the dephasing parameters, we first present the coefficients and the quantum-state structure, followed by the analytical expression for the quantum discord. 
To gain deeper insight into the system’s dynamics, we then analyze the behavior of linear-entropy-based QD by considering two distinct regimes: the symmetric case where $\gamma_A(t) = \gamma_B(t)= \gamma(t)$, and the asymmetric case where $\gamma_B(t) = 0$ while $\gamma_A(t)$ takes different values. At this point, the reduced state of qubit $A$ is expressed in the Pauli matrix basis. In particular, it takes the form
\begin{equation}
\rho^A = \frac12 \left( I_2 + R_{30} \, \sigma_3 \right) = \frac12 \begin{pmatrix} 1 + R_{30} & 0 \\ 0 & 1 - R_{30} \end{pmatrix},
\end{equation}
where $\sigma_3$ is the third Pauli matrix. The diagonal elements of $\rho^A$ give the probabilities of finding the qubit in state $\ket{0}$ or $\ket{1}$, controlled by the parameter $R_{30}$.\par
The global pure state of the hybrid system, denoted by $\ket{v^{A/A'}}$, represents the symmetric purification of the reduced density operator $\rho^A$ with an auxiliary system $A'$. It is a coherent superposition of tensor product states. We can write
\begin{equation}
\ket{v^{A/A'}} = \sqrt{\frac{1 + R_{03}}{2}} \, \ket{00} + \sqrt{\frac{1 - R_{03}}{2}} \, \ket{11},
\end{equation}
where $\ket{00}$ and $\ket{11}$ denote the tensor product states between qubit $A$ and two effective levels of qutrit $B$. This state is entangled between the qubit and the qutrit as long as $R_{03} \neq \pm 1$.\par
The bipartite density matrix reduced to an effective $2 \times 2$ subspace is written as
\begin{equation}
\rho^{A/A'} = \frac12 
\begin{pmatrix} 
1 + R_{03} & 0 & 0 & \sqrt{1 - R_{03}^2} \\ 
0 & 0 & 0 & 0 \\ 
0 & 0 & 0 & 0 \\ 
\sqrt{1 - R_{03}^2} & 0 & 0 & 1 - R_{03}
\end{pmatrix}.
\end{equation}
This matrix is Hermitian 
and describes the qubit-qutrit correlations in the considered subspace.

The coefficients \(R_{\alpha\beta}\) of the generalized Pauli decomposition mentioned in Eq.~\eqref{R} 
depend on both the diagonal elements and the off-diagonal elements of the density matrix, whose explicit expressions are presented as 
\begin{align}  \label{eq(44)}
R_{00} &= R_{33} = 1, \nonumber\\
R_{11} &= R_{22} = \sqrt{1-R_{03}^{2}},\nonumber\\
R_{30} &= \frac{2}{3} (\rho_{11} + \rho_{22} + \rho_{33} - \rho_{44} - \rho_{55} - \rho_{66}), \nonumber\\
R_{03} &= \frac{2}{3} (\rho_{11} - \rho_{22} + \rho_{44}- \rho_{55}), \nonumber\\
R_{08} &= \frac{2}{3\sqrt{3}} (\rho_{11} + \rho_{22} - 2\rho_{33} + \rho_{44} + \rho_{55} - 2\rho_{66}), \nonumber\\
R_{31} &= \frac{2}{3} \big[k(\rho_{24} + \rho_{24}^{*}) + l(\rho_{35} + \rho_{35}^{*})\big], \\ 
R_{32} &= \frac{2i}{3} \big[k(\rho_{24} - \rho_{24}^{*}) + l(\rho_{35} - \rho_{35}^{*})\big], \nonumber\\
R_{33} &= \frac{2}{3} (\rho_{11} - \rho_{22} - \rho_{44} + \rho_{55}), \nonumber\\
R_{36} &= -\frac{2}{3} l (\rho_{35} + \rho_{35}^{*}), \qquad
R_{37} = \frac{2i}{3} l (\rho_{35} - \rho_{35}^{*}), \nonumber\\
R_{38} &= \frac{2}{3\sqrt{3}} (\rho_{11} + \rho_{22} - 2\rho_{33} - \rho_{44} - \rho_{55} + 2\rho_{66}).\nonumber
\end{align}

The matrix $M$ mentioned in Eq.~\eqref{LLL} is now expressed, after substituting the expressions of the coefficients $R_{\alpha\beta}$ and simplifying, in the following form
\begin{equation}
\resizebox{\columnwidth}{!}{$
M = \frac{2}{3R_{11}^2} 
\begin{pmatrix}
0 & 0 & 0 & 0 & 0 & 0 & 0 & 0 \\
0 & 0 & 0 & 0 & 0 & 0 & 0 & 0 \\
R_{31} & R_{32} & R_{33} - R_{03}^2 & 0 & 0 & R_{36} & R_{37} & R_{38} - R_{03}R_{08}
\end{pmatrix}.
$}
\end{equation}
It can be observed that only the elements in the third row are non-zero, indicating that the matrix $M$ is of rank 1. 
Since only one eigenvalue $\lambda_1$ is non-zero, reflecting the sparse structure of $M$ and the presence of a single dominant correlation mode in the system, this simplification allows for the direct calculation of the QD using only $\lambda_1$, which is given by
\begin{equation}
\begin{aligned}
\lambda_1 = \frac{4}{9R_{11}^4} \Big[ &\, R_{31}^2 + R_{32}^2 + (R_{33} - R_{03}^2)^2 + R_{36}^2 \\
& + R_{37}^2 + (R_{38} - R_{03} R_{08})^2 \Big].
\end{aligned}
\end{equation}
Using the expression defined in Eq.~\eqref{Jl} and the coefficients $R_{\alpha\beta}$ from Eq.~\eqref{eq(44)}, we can compute the classical part of the total correlations, $J_2(\rho)$. Substituting the expression for $\lambda_{\max} = \lambda_1$ derived previously, we obtain
\begin{equation}
\begin{split}
J_2(\rho) = \frac{1 - R_{30}^2}{R_{11}^4} \Big[ & R_{31}^2 + R_{32}^2 + (R_{33} - R_{03}^2)^2 \\
 &+ R_{36}^2 + R_{37}^2 + (R_{38} - R_{03} R_{08})^2 \Big].
\end{split}
\end{equation}
This result can be equivalently expressed in terms of the maximum eigenvalue as
\begin{equation}
J_2(\rho)= \frac{4}{9} \lambda_{1} (1 - R_{30}^2).
\end{equation}
Following the same steps as for the dephasing channel, we now examine the phase-flip channel by considering the configuration defined in Eq.~\eqref{eq28}. Therefore, the QD is obtained as indicated in Eq.~\eqref{Q General} and corresponds to the difference between the mutual information \(I(\rho)\) and the classical part \(J_2(\rho)\), {where all correlation coefficients $R_{\alpha\beta}$, and consequently the matrix $M$, are evaluated after incorporating the effect of the phase-flip channel on the off-diagonal elements of the density matrix. In particular, the coherences are modified according to the transformations $\rho_{24} \rightarrow k^2 \rho_{24}$ and $\rho_{35} \rightarrow k^2 \rho_{35}$. As a result, the matrix $M$ is correspondingly altered, which leads to a modified expression for the maximal eigenvalue $\lambda_{\max}(M^T M)$, and hence affects the classical correlations $J_2(\rho_{\text{Pc}})$.}

\begin{figure*}[t]    
	\begin{minipage}[b]{0.48\linewidth}
		\centering
		\includegraphics[scale=0.376]{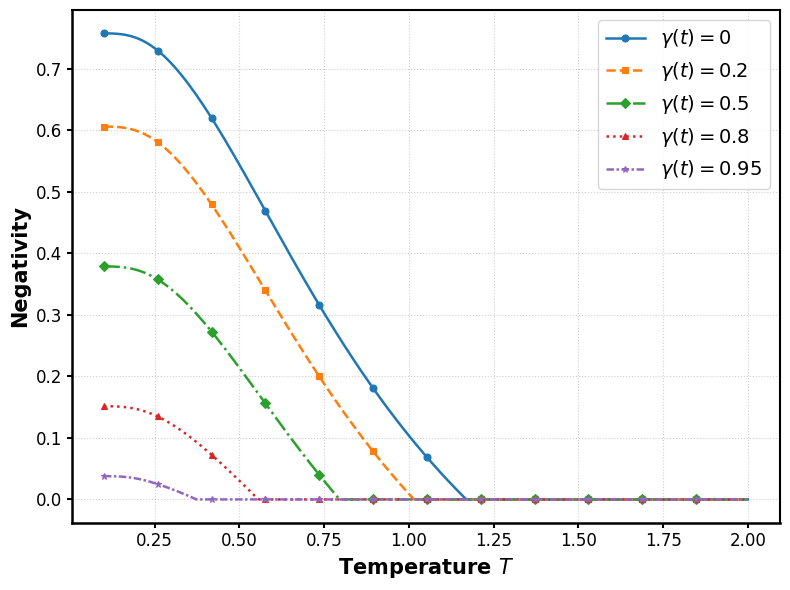}
        \put(-27,45){(a)}
	\end{minipage} \hfill
	\begin{minipage}[b]{0.48\linewidth}
    
		\centering
		\includegraphics[scale=0.376]{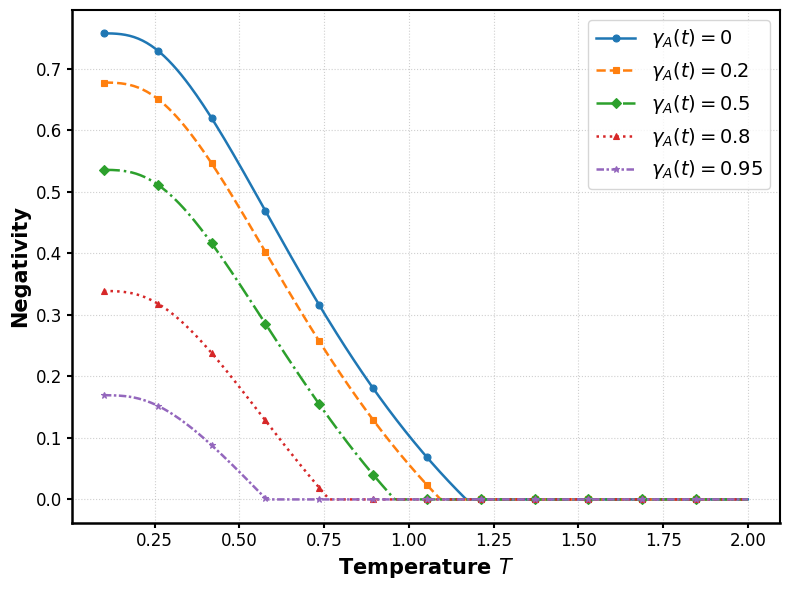}
        \put(-27,45){(b)}
	\end{minipage}

	\begin{minipage}[b]{0.48\linewidth}
		\centering
		\includegraphics[scale=0.376]{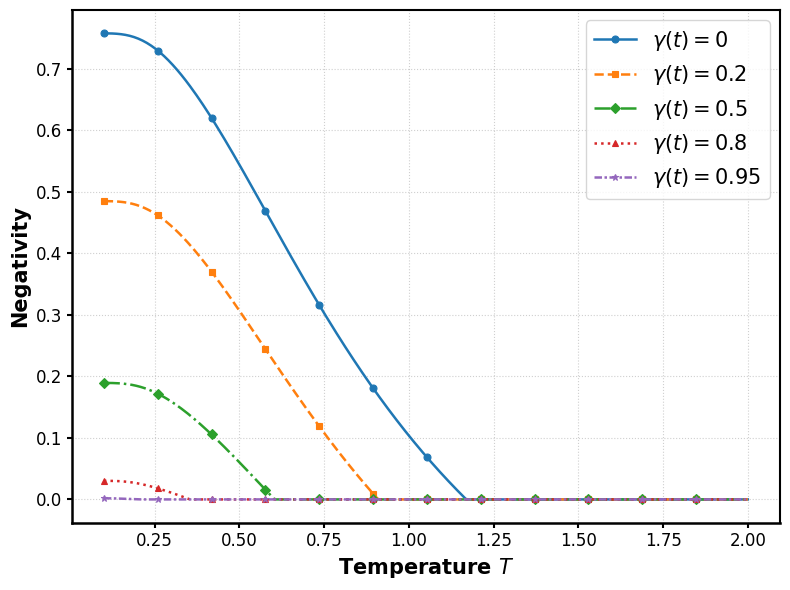}
        \put(-27,45){(c)}
	\end{minipage} \hfill
	\begin{minipage}[b]{0.48\linewidth}
    
		\centering
		\includegraphics[scale=0.376]{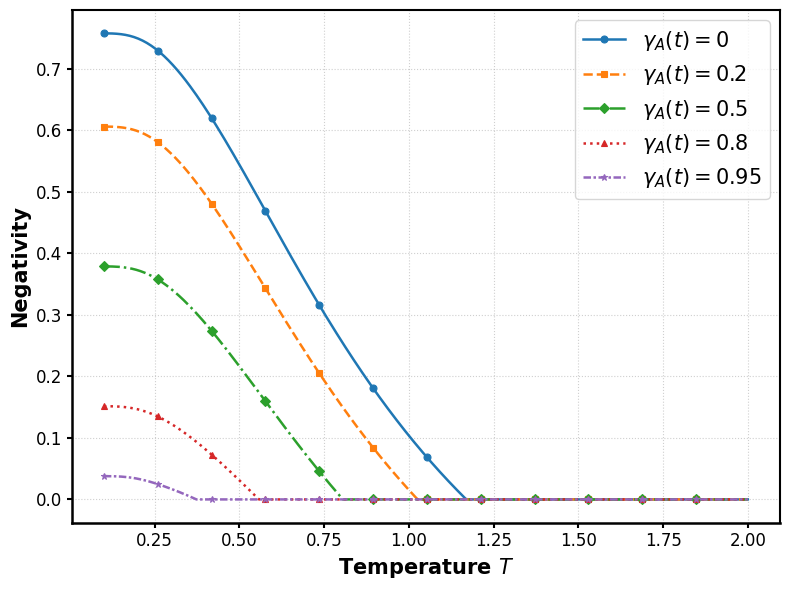}
        \put(-27,45){(d)}
	\end{minipage}

    \caption{Negativity as a function of temperature for the hybrid system under different noisy channels: (a) and (b) correspond to the dephasing channel, while (c) and (d) correspond to the phase-flip channel. Plots (a) and (c) correspond to $\gamma_A(t)=\gamma_B(t)=\gamma(t)$ and (b) and (d) correspond to $\gamma_B(t)=0$. Fixed parameters are: $J=0$, $J_z=1$, $K=0.2$, $K_1=-0.1$, $K_2=0.22$, $B_1=0.3$, $B_2=-0.7$, $D_z=0.32$, $\Gamma=-0.87$, and $\Lambda=0.31$. 
}
    \label{Negativity in chan 1 and 2}

\end{figure*}

\section{Results and Discussion} \label{Sec 6}
\subsection{Negativity as a function of temperature}
The behavior of negativity shown in Fig.~\ref{Negativity in chan 1 and 2}(a) and \ref{Negativity in chan 1 and 2}(b) provides a clear physical picture of how thermal effects and phase noise jointly degrade entanglement in a qubit–qutrit system subjected to dephasing. As the temperature increases, thermal excitations enhance classical fluctuations of the environment, which in turn accelerate the randomization of relative phases between the system’s energy eigenstates. Since dephasing directly suppresses quantum coherence without exchanging energy, this phase randomization progressively destroys the off-diagonal density-matrix elements responsible for entanglement, leading to the observed monotonic decrease of negativity with temperature.

From Fig.~\ref{Negativity in chan 1 and 2}(a), in the absence of decoherence ($\gamma_A(t)=\gamma_B(t)=\gamma(t) = 0$), one can see that the system retains a high degree of quantum coherence at low temperatures, and the negativity attains a large value $N(\rho_{\text{Dc}})\approx 0.75$, signaling strong qubit–qutrit entanglement. Introducing dephasing ($\gamma(t) > 0$) accelerates the loss of coherence, so that even moderate thermal noise becomes sufficient to fully suppress entanglement. This manifests as entanglement sudden death, where the negativity drops to zero at a finite temperature rather than vanishing asymptotically, highlighting the nontrivial interplay between temperature-induced mixedness and dephasing dynamics.

In the asymmetric regime $\gamma_B(t) = 0$ (see Fig.~\ref{Negativity in chan 1 and 2}(b)), the negativity retains slightly higher values at low temperatures, suggesting that asymmetry in the noise parameters delays the complete disappearance of quantum correlations.
{This behavior can be understood from two complementary physical mechanisms. First, when decoherence acts only on one subsystem, the overall environmental disturbance affecting the bipartite system is effectively reduced. Since dephasing primarily suppresses the off-diagonal elements of the density matrix, limiting its action to a single subsystem naturally slows down the global loss of coherence. Second, beyond this global reduction effect, the asymmetry itself plays a nontrivial role in the preservation of quantum correlations. As seen from the structure of the evolved density matrix in Eq. (\ref{eq21}), the coherence terms are modulated by factors depending on both $\gamma_A(t)$ and $\gamma_B(t)$. In the asymmetric case $\gamma_B(t)=0$, part of these coherence contributions remains only partially degraded, allowing the unaffected subsystem to act as a partial reservoir of coherence. As a result, the destruction of entanglement is less cooperative than in the symmetric case, leading to a slight delay in the disappearance of negativity. Therefore, the enhanced robustness observed in Fig.~\ref{Negativity in chan 1 and 2}(b) originates from the combined effect of reduced effective noise strength and the asymmetric distribution of decoherence.}

Comparing the symmetric ($\gamma_A(t)=\gamma_B(t)$) and asymmetric ($\gamma_B(t)=0$) dephasing regimes further clarifies the role of noise distribution. When both subsystems are equally affected, phase noise acts cooperatively to destroy inter-system coherence, resulting in a faster decay of entanglement. In contrast, when dephasing acts only on one subsystem, part of the quantum coherence is effectively shielded, allowing the negativity to survive up to slightly higher temperatures. This indicates that asymmetry in environmental coupling can partially protect quantum correlations, delaying their complete disappearance even though the overall trend of entanglement degradation with temperature remains unavoidable.

The plots in Fig.~\ref{Negativity in chan 1 and 2}(c) and \ref{Negativity in chan 1 and 2}(d) represent the behavior of negativity for the phase-flip channel. The general behavior remains similar, but the decay is faster. 
Regarding Fig.~\ref{Negativity in chan 1 and 2}(c), for $\gamma(t) = 0$, the entanglement already vanishes at $T \approx 1.2$, and for higher values of $\gamma(t)$ ($0.2$, $0.5$, $0.8$, $0.95$), the negativity becomes zero from $T \approx 0.9$. This result shows that the phase-flip type channel induces stronger decoherence than the dephasing channel. 
In the asymmetric regime [Fig.~\ref{Negativity in chan 1 and 2}(d)], the negativity exhibits a slightly more pronounced resistance at low temperature, although the general trend of rapid entanglement disappearance remains unchanged.

Fundamentally, the difference in behavior between the two channels is explained by the nature of the coupling to the environment: the dephasing channel perturbs the coherence of the state amplitudes, while the phase-flip channel directly affects the relative phase between the system's energy levels, leading to faster degradation of quantum correlations.
\begin{figure*}[t]    
	\begin{minipage}[b]{0.48\linewidth}
		\centering
		\includegraphics[scale=0.376]{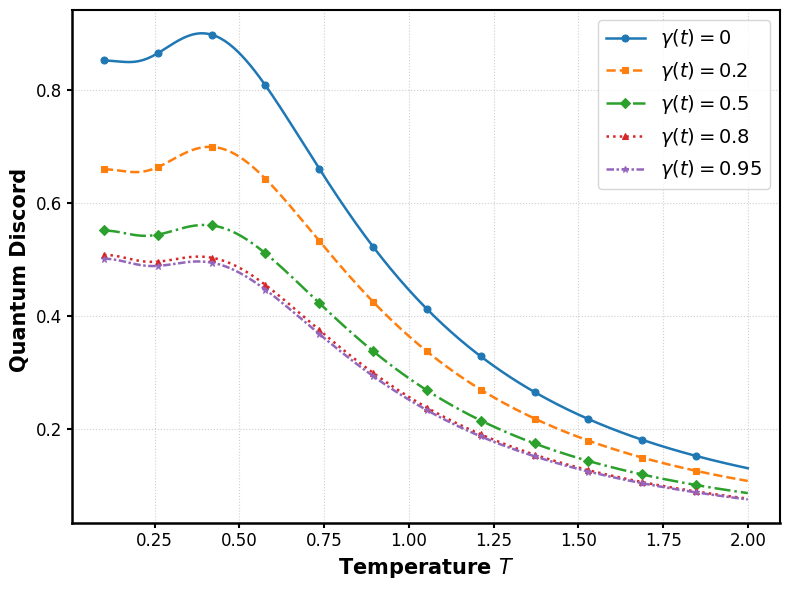}
        \put(-27,45){(a)}
	\end{minipage} \hfill
	\begin{minipage}[b]{0.48\linewidth}
    
		\centering
		\includegraphics[scale=0.376]{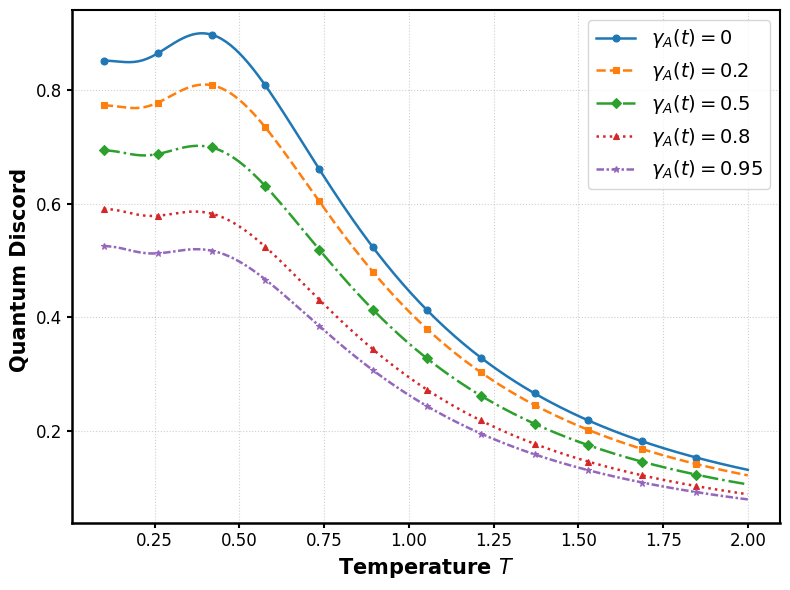}
        \put(-27,45){(b)}
	\end{minipage}

	\begin{minipage}[b]{0.48\linewidth}
		\centering
		\includegraphics[scale=0.376]{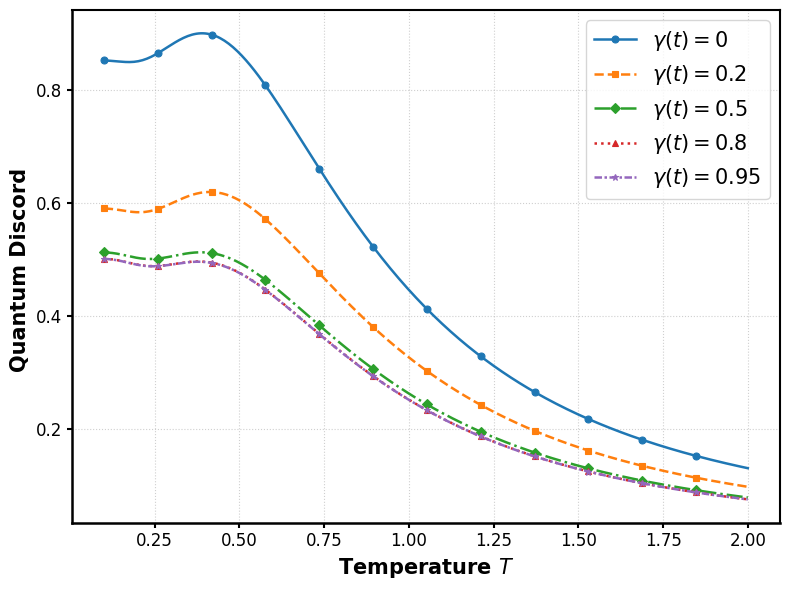}
        \put(-27,45){(c)}
	\end{minipage} \hfill
	\begin{minipage}[b]{0.48\linewidth}
    
		\centering
		\includegraphics[scale=0.376]{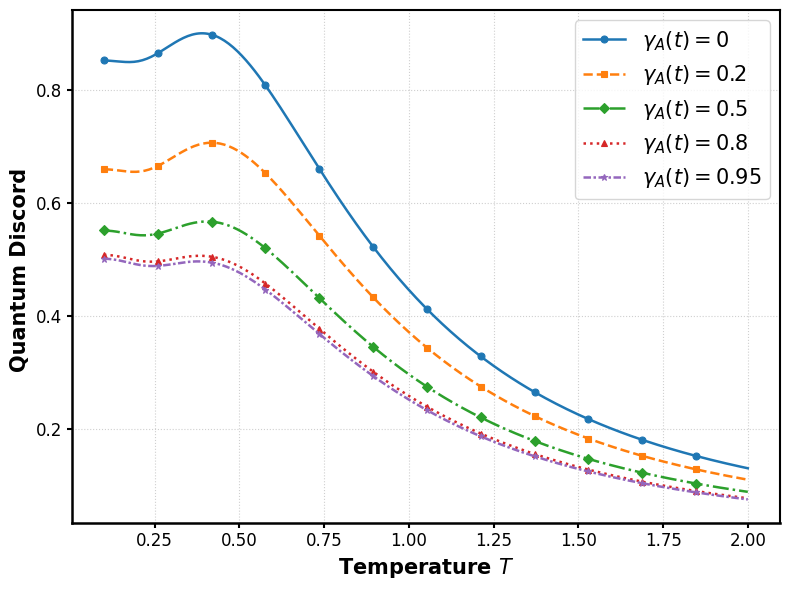}
        \put(-27,45){(d)}
	\end{minipage}
    \caption{Similar to Fig.~\ref{Negativity in chan 1 and 2} but for linear-entropy-based QD.} 
    \label{QD based on Linear entropy Fig}

\end{figure*}
\subsection{Linear-entropy-based QD versus temperature}
The curves in Fig.~\ref{QD based on Linear entropy Fig}(a) illustrate the evolution of QD as a function of temperature for different values of the dephasing parameter $\gamma(t)$. QD, formulated from linear entropy, quantifies the total quantum correlations, including those that persist when entanglement, measured by negativity, vanishes. It is observed that QD globally decreases with temperature, reflecting the progressive degradation of quantum correlations under the effect of thermal mixing. { However, a noticeable local maximum appears in the low-temperature region ($T \approx 0.3\text{--}0.5$), revealing a phenomenon of transient thermal activation of quantum correlations. This behavior originates from the thermal redistribution of populations among the eigenstates of the system. At very low temperatures, the system is almost frozen in its ground state. A slight increase in temperature allows partial occupation of excited states that possess higher quantum coherence and stronger inter-level superpositions. This thermal population transfer temporarily enhances the amount of quantum correlations before the dominant thermal fluctuations at higher temperature progressively destroy them. It is important to note that this transient maximum is clearly observed for weak and moderate dephasing, whereas for larger values of the dephasing parameter ($\gamma=0.8$ and $0.95$) the maximum of QD occurs at the very beginning of the temperature scale ($T \to 0$). In this strong decoherence regime, the rapid suppression of coherence prevents the thermal activation from manifesting in the interval ($T \approx 0.3\text{--}0.5$).}
Increasing the dephasing parameter $\gamma$ accentuates this decay, confirming the sensitivity of quantum phase coherence to decoherence induced by the dephasing channel. 

For Fig.~\ref{QD based on Linear entropy Fig}(b), when the phase noise acts asymmetrically, i.e., on only one of the two subsystems, the decrease in QD remains slower, and the QD values stay higher across all temperatures. This asymmetry grants the system increased robustness, as the non-dephased sub-part acts as a reservoir of coherence, partially maintaining the inter-subsystem correlations. 

This behavior indicates that the non-uniform distribution of noise between the qubit and the qutrit prolongs the lifetime of quantum correlations, even in a dissipative environment.

A similar overall tendency is observed for the phase-flip channel, as illustrated in Fig.~\ref{QD based on Linear entropy Fig}(c) and Fig.~\ref{QD based on Linear entropy Fig}(d). The QD decreases monotonically with temperature, highlighting the detrimental effect of phase-flip noise on quantum correlations. { Nevertheless, the same transient thermal activation at low temperature can be identified here, confirming that this effect is not noise-type dependent but rather linked to the intrinsic structure of the thermal state of the system.} Compared to the dephasing case, the decay of QD under phase-flip noise is slightly more pronounced, indicating a stronger loss of coherence under this type of noise. The symmetric configuration $\gamma_A(t) = \gamma_B(t)$ leads to a faster suppression of correlations, while the asymmetric case $\gamma_B(t) = 0$ again demonstrates greater robustness, preserving nonclassical correlations over a broader temperature range.

\subsection{Comparative analysis }
{In the previous two subsections, we have compared symmetric and asymmetric noise configurations across both subsystems to understand their impact on quantum correlations. In the symmetric case, both subsystems experience identical decoherence $\gamma_A(t) = \gamma_B(t)$, which leads to a cooperative degradation of quantum correlations, causing them to decay more rapidly. In contrast, in the asymmetric case, only one subsystem (the qubit) is affected by noise $\gamma_B(t) = 0$, which allows part of the quantum coherence to be preserved in the unaffected subsystem. This results in enhanced robustness of both entanglement and QD, as clearly shown in Figs. \ref{Negativity in chan 1 and 2} and \ref{QD based on Linear entropy Fig}, where correlations persist over a wider range of temperatures. This behavior highlights the protective role of non-uniform environmental coupling in hybrid qubit-qutrit systems and provides deeper insight into the interplay between decoherence distribution and correlation dynamics.

Note that the situation where the roles of the subsystems are inverted, namely $\gamma_A(t)=0$ and $\gamma_B(t)>0$, can be naturally addressed within the same theoretical framework.
We have explicitly analyzed this configuration and found that it leads to qualitatively similar behavior of both entanglement and quantum correlations compared to the case where $\gamma_B(t)=0$ and $\gamma_A(t)>0$. In particular, the overall trends of degradation, robustness under asymmetric noise, and the comparative effects of the different channels remain unchanged, with only minor quantitative differences.
As a result, exchanging the roles of the qubit and the qutrit does not introduce fundamentally new physical features.
Therefore, in order to avoid redundancy and to maintain a clear and concise presentation, we have chosen to display only one asymmetric configuration in the manuscript, which is sufficient to capture the essential physics discussed in this work.}

\subsection{Final remarks}
{It is worth emphasizing that the enhanced robustness of quantum correlations observed under asymmetric noise is not specific to the $2 \otimes 3$ system considered here, but rather reflects a more general feature of uneven environmental coupling. When decoherence acts only on one subsystem, part of the global coherence can be preserved in the unaffected subsystem, which slows down the degradation of both entanglement and quantum discord. Nevertheless, the hybrid dimensionality plays a non-negligible quantitative role. In particular, the qutrit subsystem, due to its larger Hilbert space, supports a richer structure of coherence and correlations, which can contribute to a slightly improved resistance against decoherence compared to the qubit. Therefore, while the qualitative behavior remains general, the specific robustness observed in our results is partially influenced by the $2 \otimes 3$ dimensional structure of the system.
}

\section{Summary} \label{Sec 7}
In this work, we have investigated the thermal and decoherence-induced dynamics of quantum correlations in an axially symmetric qubit–qutrit system governed by a physically motivated Hamiltonian. By analyzing both entanglement, quantified through negativity, and more general quantum correlations described by quantum discord, we have provided a comprehensive picture of how environmental effects influence hybrid quantum systems.

Our results show that entanglement is highly sensitive to temperature and noise, exhibiting a monotonic decay with increasing temperature and decoherence strength. In particular, the phase-flip channel is found to be more detrimental than the dephasing channel, leading to faster entanglement suppression and earlier onset of entanglement sudden death. Moreover, the asymmetric noise configuration, in which only one subsystem is affected, slightly delays entanglement disappearance, highlighting the role of noise distribution in controlling quantum correlations.

Beyond entanglement, the analysis of quantum discord reveals a richer and more resilient behavior. While quantum discord also decreases with temperature due to thermal mixing and decoherence-induced loss of coherence, it remains nonzero even after entanglement vanishes. Notably, unlike negativity, the temperature dependence of quantum discord exhibits a local maximum at low temperatures, indicating a transient enhancement of quantum correlations induced by thermal activation before their eventual decay. This feature represents a clear qualitative difference between entanglement and discord and underscores the fundamentally broader nature of quantum correlations captured by quantum discord. Furthermore, asymmetric noise configurations consistently preserve higher levels of discord, demonstrating that partial protection of coherence can significantly extend the lifetime of quantum correlations.

\vspace{0.5cm}

\noindent \textbf{Data availability:} All data generated or analyzed during this study are included in this published article.
\\
\\
\noindent \textbf{Competing interests:} The authors declare no competing interests.
\\
\\
\noindent \textbf{Author contributions:} M.A., N.E.A., A.S., and R.A.L. have contributed to the writing of the manuscript and
interpretation of the results. A.S. and S.H. performed thorough reviews of the manuscript and confirmed the conclusions.
\\
\\
\noindent \textbf{Funding statement:}
This research did not receive a specific grant from any funding agency in the public, commercial or non-profit sectors.

\end{document}